\documentstyle[12pt]{article}
\begin{document}
\title{Replication and Evolution of Quantum Species}
\author{Arun K. Pati $^{1}$ \\
Institute of Physics, Sainik School Post,\\
Bhubaneswar-751005, India \\                                
and \\
$^{1}$ School of Informatics, \\
University of Wales, Bangor LL 57 1UT, United Kingdom}

\def\ra{\rangle} \def\la{\langle} \def\ver{\arrowvert}

\maketitle

\begin{abstract}
We dwell upon the physicist's conception of `life' since
Schr{\"o}dinger and Wigner through to the modern-day language of
living systems in the light of quantum information. We discuss some
basic features of a living system such as ordinary replication and
evolution in terms of quantum bio-information.  We also discuss the
principle of no-culling of living replicas. We show that in a
collection of identical species there can be no entanglement between
one of the mutated copies and the rest of the species in a closed
universe. Even though these discussions revolve around `artificial
life' they may still be applicable in real biological systems under
suitable conditions.
\end{abstract}


\section{Introduction}
The desire to understand life in a dynamical manner is as old as modern
civilization \cite{wme,sch,pcwd}. The question that has bothered some of the
best minds of the
last several centuries is: Can the dynamics of living systems be truly
different to that of the inanimate objects? The later objects in the
macroscopic
world are described by Newtonian laws of motion
which are completely deterministic.
Although inanimate systems have order at the macroscopic level, they may
nevertheless be disordered at the microscopic level.
However, all  living systems display
orderliness from the cell level through to the level of DNA.
As Schr{\"o}dinger has beautifully paraphrased in his classic work:
{\it  ``In biology we are faced with an entirely different situation.
A single group
of atoms existing only in one copy produces orderly events, marvelously
tuned in with each other and with the environment according to most subtle
laws''} \cite{sch}.  Order in a system can arise in two ways, one possibility
is `order
from disorder' and the other is `order from order'. How can such orderly
events arise in  living systems?
Many would support the view that life emerges by `order from chaos'
and many others support that `order comes from order'. Schr{\"o}dinger had
firmly believed that the behavior of living systems is based on the
`order-from-order' principle.

But what mechanics governs living systems? Does it have to be classical
mechanics or the ultimate theory of nature such as quantum
mechanics? No one knows for sure, because no one knows what the word
`life' really means. There have been many attempts to define life in a
mathematical
abstract manner \cite{gjc}. But it is evidently clear that there is some
special quality of a living system, which is called `life'
\cite{sch,pcwd,gjc,ca}.
All we know is certain features of living systems that distinguish them
from non-living objects. For example, there is no doubt that a living
system can replicate, evolve, and moreover it can self-reproduce, self-repair
and so on, which are not generally seen in non-living systems.  In non-living
systems one can observe phenomena of replication and evolution. Replication
of non-living objects (such as a piece of paper containing some information)
or a living object (such as a cell) is a process where an input is fed
through a suitable device and at the output one obtains two identical copies.
Evolution of non-living or living systems involves dynamical
interaction where the initial structure can evolve to a new structure 
consistent with physical laws.

There is growing thought among physicists \cite{sch,penrose}
and biologists \cite{fad1} that quantum mechanics might play an important
role in living systems. McFadden has the view that ultimately the origin
of life will be answered by quantum mechanics \cite{fad1}.
To cite a few others, Frolich has suggested that the action of enzymes in
biological systems would be understood from quantum mechanical principles
\cite{fro}. Home and Chattopadhyaya have
proposed a way to describe quantum mechanical measurement process
in a DNA molecule,
which may be in a superposition of mutational states \cite{dc}.
McFadden and Al-Khalili have argued that during cell mutation quantum
mechanics may play an important role and entanglement between the mutational
state and the environment may enhance the probability of mutation \cite{fad}.
Patel has proposed that a quantum algorithm may be at work in genetic evolution
\cite{patel}, though skepticism prevails on how
quantum coherence can be maintained in a DNA molecule.

Notably, Wigner was baffled by the fact that ``{\it there are
organisms with certain structures which, if brought into contact with
certain nutrient materials multiply in order to produce identical structures
to that of the original one''} \cite{wigner}. So he addressed the following
question:
Can laws of physics describe living matter? Especially, he had first
asked the question of
replication of quantum states, which in the language of quantum information
is called `cloning' of quantum states \footnote{
Though the title of Wigner's paper  in \cite{wigner}
says that it addresses the question of a `self-reproducing' unit, it actually
addresses the question of `reproducing' unit only.
}. However,
he could not come to a doctrine such as the `no-cloning'
principle \cite{wz,dd}.
He purports to argue that in quantum world it is infinitely unlikely that
such a replicating
machine could exist \cite{wigner}. It may be remarked that the process of
replication and  self-replication are different games
altogether. In the replication process any input is supposed to be
duplicated ---
whereas, in self-replication, the state along with its program would be
duplicated. That is, by furnishing a suitable description of the program,
the machine will construct a copy of the input species as well as a copy
of the instruction. In replication or cloning, one does not need to produce
a copy of the program. The question of self-replication in a mechanical
context was asked by von Neumann \cite{von}. He had explicitly shown that
in the classical world there exists a machine called a `universal
constructor'. If it is provided with its complete specification
then it can self-reproduce. Fifty years later, it was asked if one
can design a quantum mechanical universal constructor that can self-reproduce
any arbitrary state along with its program. It is found that there cannot
exist an all-purpose universal quantum constructor in a closed universe with
finite resource \cite{arun}.
If one studies emergence of life from a point of view of game theory
then interesting quantum effects can be incorporated. Flitney and
Abbott have recently studied Conway's game of life in a semi-quantum mechanical
context \cite{fd}.

In this paper, we discuss various features of a living system
such as the process of replication, culling of identical replicas, and
entanglement between mutated and the rest of the replicas
from quantum information point of view.
In Section 2, we discuss the original version of Wigner's replication machine. In
Section 3, we discuss how Wigner's replication process could be ruled out
simply using the linearity or unitarity of quantum theory. Also we will mention
how one can have replication of quantum species states in a stochastic manner
with a certain probability of success. In Section 4, we discuss the
no-culling principle for living replicas. In Section 5, we will discuss
a paradox of how entangling several copies of a living organism
with one of its mutated versions nullifies the interaction. Then, we will
show why it is not possible to entangle one copy with a mutated copy
of a living state. Then our conclusion follows in the Section 6.
It may be mentioned that we are not talking of `real' biological systems
in their full detail but rather some kind of `artificial' living system
in the quantum mechanical sense. The main motivation here is to discuss
and contrast this virtual `world of quantum mechanical life' with
real biology. Though some ideas may still apply to real biological
systems, whether one can apply all these ideas to real bio-systems
will depend on various other factors, such as decoherence time at the
molecular level.

\section{Wigner's Replication Machine}

In the discussion of replicating machines Wigner assumed that there be a
 `living state' $\ver \psi \ra$ which is given in a quantum
mechanical sense in a finite dimensional Hilbert space ${\cal H}^N$.
This may store some `bio-information' which may undergo the replication
process. Further, there exists at least one nutrient state
$\ver w \ra \in {\cal H}^{NR}$ which will allow the organism to replicate
onto it. Therefore, the
initial state of the system, namely, the organism and nutrient is given by

\begin{equation}
\ver \Psi_i\ra = \ver \psi \ra  \ver w \ra.
\end{equation}
After the replication process the combined state would be given by
\begin{equation}
\ver \Psi_i \ra \rightarrow \ver \Psi_f \ra = U \ver \Psi_i \ra =
\ver \psi \ra  \ver \psi \ra  \ver r \ra,
\end{equation}
where two organisms have been produced in the same state $\ver \psi \ra$
and $\ver r \ra$ would be the rejected part of the nutrient state that may
belong to a Hilbert space of dimension $R$. The interaction is unitary, which
means that there is some Hamiltonian underlying this process. It was
assumed that the Hamiltonian that governs the behavior of such a complex
organism would be a random symmetric matrix since nothing much can be known
about it. If such a unitary interaction
occurs between the organism and the nutrient, then in terms of its components
(with respect to some chosen basis) in their respective Hilbert spaces,
the replication process can be described by

\begin{equation}
\psi_k  \psi_l  r_{\mu}  =\sum_{k'l'\mu'} \la k l \mu |U| k'l'\mu'\ra
 \psi_{k'}  w_{l'\mu'} .
\end{equation}

The question Wigner posed is: Given an interaction, described by a unitary
operator $U$, is it possible
to find $N$ numbers $\psi_k $, which together with suitably chosen
$R$ numbers $ r_{\mu}$ and $NR$ numbers of $ w_{l\mu} $
that satisfies (3)? He argued that since (3) must be valid for any $k, l$ and
$\mu$, there are a total of $2N^2R$ real equations. On the other hand there are only
$2(N+R +NR)$ real unknowns. Since the number of unknowns are very much fewer
than the number of equations, proclaimed in his words,
{\em  it would be a miracle if (3) would be satisfied.}
This was Wigner's reasoning that it is infinitely unlikely that there can be
a replicating machine for a living organism in the quantum mechanical sense.

In his thought provoking article, he went on providing further arguments to
support his claim. However, he assumed something erroneous in between. He
said that if $\ver k \ra$ represents a `living state', then any linear
superposition of such basis living states would also represent a living state.
This means that a state of the type $\ver \psi \ra= \sum_k \psi_k
\ver k\ra$ would also be a replicating state. But now we know that
it is precisely this linearity of superposition that prohibits the replication of
a single quantum state \cite{wz,dd}.

\section{Replication of Quantum Species}

In the language of quantum information theory, it is now well known that an
arbitrary quantum state cannot be replicated---called the
`no-cloning' principle. This is due to the seminal work of
Wootters-Zurek \cite{wz} and Dieks \cite{dd}.
Let us see how Wigner's replicating machine for a species can be
ruled out simply based on the linearity of the quantum theory.
Here a quantum species and organism will be used in the same sense  which
means possibly a smallest unit of a living system that respresents either
`artifical' or `real' life.

The question is can there be a process such that the
initial state of the organism and nutrient given by

\begin{equation}
\ver \Psi_i\ra = \ver \psi \ra \ver w \ra
\end{equation}
can be transformed to

\begin{equation}
\ver \Psi_f \ra= \ver \psi \ra \ver \psi \ra  \ver r \ra,
\end{equation}
where $\ver r \ra$ is rejected part of the nutrient state?

Suppose that there is a process where one can replicate orthogonal
basis states

\begin{equation}
\ver k \ra  \ver w \ra \rightarrow
\ver k \ra  \ver k \ra  \ver r_k \ra,
\end{equation}
where $\ver r_k \ra$ is the final nutrient state when the living state
is $\ver k \ra$.
If we take an arbitrary state $\ver \psi \ra= \sum_k \psi_k
\ver k\ra$ which is a linear superposition of orthogonal states,
then by the linearity of quantum theory we will have

\begin{equation}
\ver \psi \ra  \ver w \ra =
\sum_k \psi_k \ver k \ra  \ver w\ra \rightarrow \sum_k \psi_k
\ver k \ra  \ver k \ra  \ver r_k\ra.
\end{equation}
However, the actual replication of a living organism would be given by

\begin{equation}
\ver \psi \ra  \ver \psi \ra  \ver r \ra=
\sum_{kl} \psi_k \psi_l \ver k \ra  \ver l \ra 
\ver r \ra.
\end{equation}

Under no circumstance can (7) and (8) be identical. The simplest reason is,
in general, that (7) is an entangled state of the original and the 
nutrient state,
whereas (8) is not an entangled state. It can be seen from (7) that the
individual state of the original or the replica is a mixed state
(after tracing out the
other two subsystems) whereas in the ideal case the original or
the replica is a pure state. Therefore, we have failed to replicate an
arbitrary living system in a quantum mechanical sense.

Also one cannot replicate any two non-orthogonal
states by a unitary process \cite{hy}.
Non-orthogonality of two states physically would mean that there is a finite
overlap between them, i.e., they are not completely distinct. In the biological
world this would probably mean two species described by non-orthogonal state
have some similarity in their information content, behavior, or
qualities. Let us see how to rule out Wigner's replicating machine for
two non-orthogonal states by a physical interaction (whether it is described
by a random Hamiltonian or not does not really matter). Suppose it is possible
to satisfy the following two equations for two species represented by
two non-orthogonal states:

\begin{eqnarray}
\ver \psi_1 \ra  \ver w \ra &\rightarrow&
\ver \psi_1 \ra  \ver \psi_1 \ra  \ver r_1 \ra \nonumber\\
\ver \psi_2 \ra  \ver w \ra &\rightarrow&
\ver \psi_2 \ra  \ver \psi_2 \ra  \ver r_2 \ra.
\end{eqnarray}
Then it must preserve the inner product. This implies that we have

\begin{equation}
\la \psi_1 \ver \psi_2 \ra = \la \psi_1 \ver \psi_2 \ra^2 \la r_1 \ver r_2 \ra.
\end{equation}
But this can never be satisfied by a unitary process. This is so, because (10)
implies that $|\la \psi_1 \ver \psi_2 \ra| \le |\la \psi_1
\ver \psi_2 \ra|^2$, (using
the condition that  $|\la r_1 \ver r_2 \ra| \le 1$) which implies
a contradiction. However, for two orthogonal species states (10) can
hold, since we have $\la \psi_1 \ver \psi_2 \ra =0$. This is consistent with
the understanding that we store classical information in the orthogonal states
and the former can be perfectly copied so also the later.

Schr{\"o}dinger, Wigner and probably others too have suspected that quantum
mechanics would not permit the accurate replication of biological information.
We now understand that it is indeed true. However, biological organisms do
replicate information. The explanation for this might therefore be because
(i) they transcend the laws of quantum world or (ii) biological
systems operate in the classical regime, where accurate information
replication is possible. The one that will strike most readers is
the second explanation. This may be an appealing reason why we can clone a
sheep or a pig (a macroscopic living
object) but we cannot clone the fundamental building blocks that
constitute them (such as an electron or proton) \cite{whz}.
If a living organism is macroscopically large and distinguishable then
it is like a classical state, and hence one can clone it perfectly. In this
sense the quantum mechanical living states which have undergone
decoherence due to interaction with the environment are useful for
making perfect clones.

And if there is some error during replication then that will be very
small. For example, for an {\it E.~coli} cell the replication error is so low
that, on average an error is made only once every thousand generations.
One may ask whether this violates the no-cloning principle? The answer would be
no, because if it really violates no-cloning then that would be seen in one
generation, i.e., there should be error in each generation.
However, if quantum superposition does prevail at the
level of living organism, and they undergo replication process, then there is
something beyond linearity of quantum mechanics that may be at the root. That
would mean a modification of quantum mechanics to take into account
nonlinearity and even the process of superluminal signaling in the realm of
biological systems.

Another possibility for replication of living organisms would be through
a cyclic evolution of the state. In the quantum world, if the initial state of
a system is $\ver \psi(0)\ra= \ver k \ra$ then it can evolve into a linear
superposition of basis states such as $\sum_k \psi_k(t) \ver k\ra$
and after a certain time period
$t=T$ the state can come back to its original state, i.e. $\ver \psi(T)\ra
= \ver \psi(0)\ra =\ver k \ra$, up to an overall phase. Suppose, the
useful information (which is classical) is stored in the state $\ver k \ra$,
then the state can be copied at times $t=0, T, 2T\ldots $ and so on. However,
at any time $0 < t <T$ the state is in an arbitrary superposition and hence
cannot be replicated. In the same spirit,
if the living organism would be in a quantum
superposition of various possible states at intermediate times then
it cannot undergo replication, but it can do so periodically.
This is solely possible due to cyclic evolution of information from being
`classical' to `quantum' and back.

It is now clear that no unitary transformation can accurately replicate 
non-orthogonal quantum states. So can non-unitary transformations such 
as the measurement process help in the replication process?
The process which is ruled out by linearity alone is replication of
linearly dependent living states (i.e. if we could clone $\ver k \ra$
then we could not have cloned $\ver \psi \ra= \sum_k \psi_k \ver k \ra$).
Linear transformations include unitary as well as measurement operations.
So an arbitrary state cannot be cloned even by invoking a non-unitary
transformation. But replication is possible for linearly independent
non-orthogonal species states $\{\ver \psi_i \ra \}$ with a certain probability
of success \cite{dg1}. Therefore, by invoking
unitary and measurement processes together, it is possible to generate replicas
in a probabilistic manner.  The
probabilistic replicator of Wigner, in the light of Duan-Guo \cite{dg1} will
consist of the original, nutrient and a measuring apparatus. The transformation
would be given by

 \begin{eqnarray}
\ver \psi_i \ra  \ver w \ra  \ver P_0 \ra \rightarrow
\sqrt{p_i} \ver \psi_i \ra  \ver \psi_i \ra 
 \ver r_i \ra  \ver P_1 \ra
+  \sqrt{1-p_i} \ver \Phi_i \ra,
\end{eqnarray}
where $p_i$ is the probability of success that the replicator works,
$\ver P_0\ra$ and $\ver P_1 \ra$ are the initial and final probe
states, and $\ver \Phi_i \ra$ is some junk state of the combined living
organism, nutrient and probe. Ideally, here, a junk state would mean a
state that has no information about the original species.
By performing a measurement on the probe state the rhs of (11) undergoes
a collapse process. If the outcome is $\ver P_1 \ra$, then there are two
replicas with a
probability $p_i$. However, if the outcome is different than $\ver P_1 \ra$
then it is a failure.
The success probability of replication process of two
non-orthogonal species $\ver \psi_i \ra$ and $\ver \psi_j \ra$ is bounded by

\begin{equation}
p_{ij}=\frac{1}{2}(p_i+p_j)  \le \frac{1 }
{1+|\la \psi_i\ver \psi_j \ra|}.
\end{equation}

Obviously, this shows that the probability of replication cannot be a hundred
percent. It is even possible to create a linear superposition of a different
number of replicas of living states by similar stochastic processes
\cite{akp} of which transformation (11) may be a special case.

\section{Culling of Living Replicas}

Suppose we have a collection of replicas of a living organism. Can there be
a process which can remove a few of the replicas? i.e. take away the
information content and make the living states some nutrient states for
other species in the universe. Similar to the no-cloning ordinance in the
quantum world, we have another principle called `no-deletion' \cite{pb}.
This tells us that if we are given two or more copies of a quantum object
then it is impossible to delete the information of one copy by a physical
transformation. Here the meaning of the word `deletion' of a copy from two
copies means desiging a physical operation that transforms one of the copies to
a blank state completely (analogous to a blank paper) which has no original
information at all. In living systems, this would mean
`culling' of replicas of a organism, which must be impossible. This would
be called a `no-culling' principle in the quantum biological world
\footnote{In a private communication J.~P.~Dowling has suggested to
S.~L.~Braunstein that the `no-deletion principle' may be called `no-culling
principle'. After recalling this, now, I feel this term may suit better
for quantum mechanical living organisms.}.
Let us imagine that
we have two copies of a living organism
$\ver \psi \ra \ver \psi \ra$. The culling process will take
two copies of the organism and an ancillary state
and transform them as

\begin{eqnarray}
\ver \psi \ra \ver \psi \ra \ver r \ra \rightarrow
\ver \psi \ra \ver w \ra.
\end{eqnarray}

However, by linearity of quantum mechanical evolution one can show that
the above process cannot exist. Consider the culling process of orthogonal
living states

\begin{eqnarray}
\ver k \ra \ver k \ra \ver r \ra \rightarrow
\ver k \ra \ver w_k \ra.
\end{eqnarray}

If we have two copies of an arbitrary species $\ver \psi \ra=
\sum_{k} \psi_k \ver k \ra$ then by the linearity of quantum
theory we have

\begin{eqnarray}
\ver \psi \ra \ver \psi \ra \ver r \ra & =&
\sum_{kl} \psi_k \psi_l \ver k \ra \ver l \ra
\ver r \ra \nonumber\\
&=& \sum_{k} \psi_k^2 \ver k \ra \ver k \ra
\ver r \ra + \sum_{k\not=l} \psi_k \psi_l \ver k \ra 
\ver l \ra \ver r \ra
\nonumber\\
&\rightarrow&
\sum_k \psi_k^2 \ver k \ra \ver w_k \ra
+ \sum_{k\not=l} \psi_k \psi_l \ver \Phi_k \ra.
\end{eqnarray}

However, the resulting state given in (15) is not the one that we desire
in the culling process which is given in rhs of (13). In rhs of (13) we have
$\sum_k \psi_k \ver k \ra \ver w \ra$. Hence, one cannot perform
culling of living replicas by a linear operation. The only option is to
move the copy to the ancillary state without culling it perfectly.

The `no-cloning' and `no-culling' theorems are based on the linearity of
quantum mechanical evolution. Also based on unitarity only one can show that
it is impossible to cull a copy from two copies of non-orthogonal states 
\cite{pb1}. Recently, Jozsa has proved a much stronger
result known as the `stronger no-cloning' theorem, based on full unitarity
of the evolution of system and ancilla \cite{rj}.
A simple statement is as follows: If $\{ \ver \psi_k \ra \}$ is a
finite collection of pure states, containing no orthogonal pairs of states
and $\ver a_k \ra$ be any other set of states, then a physical operation
taking

\begin{eqnarray}
\ver \psi_k \ra \ver a_k \ra \rightarrow
\ver \psi_k \ra \ver \psi_k \ra
\end{eqnarray}
is possible if and only if there is a physical operation taking
\begin{eqnarray}
\ver a_k \ra \rightarrow \ver \psi_k \ra.
\end{eqnarray}
That is to say, in order to clone a quantum state one needs to supply
full information in the ancillary system. In his proof, the `other set of
states' could be a mixed state also. But for general readers we have stated
a simplified form of his theorem. Similarly, he has given a simple
proof of a  quantum no-deletion principle for non-orthogonal states.

As Jozsa has emphasized, to make
a replica one must supply a clone from somewhere and to delete a replica
one must move the information to somewhere. Hence, the `no-replicating' and
`no-culling' principles tell us that there is something robust about quantum
information. We can as well say that there is some permanence associated with
quantum information stored in a `artificial life' system. In the light of
these thoughts, it is very important
to realise that in a living organism we do not yet understand how the
hereditary information is passed from one generation to another without
being altered. As Schr{\"o}dinger has put, there is some permanence in
the hereditary information \cite{sch}. It may happen that if the hereditary
information is actually quantum mechanical in nature then the `no-culling'
principle may be at work in explaining some of these features. This may strike
many readers as something extraordinary, since the decoherence time scale
is far less than one second whereas evolutionary time scales are measured in
thousand of generations or more. A possible resolution may be that nature knows
how to protect the coherence of hereditary information by encoding it in
a larger Hilbert space in a sequence of time
steps by some means of self-error correcting codes, so that `no-culling'
principle works at each time step (less than the decoherence time) and
its accumulated effect is seen from generation to generation. However, this
will again depend on how well the information is protected from strong
decoherence associated with complex biological systems open to their
environment. This will deserve a separate study on its own and we do not
intend to do that here.

In real biological evolution, culling of a macroscopic species is a
classical process that takes place in an open system in which decoherence is
very strong and rapid. This is again consistent with the fact that culling
of classical information is possible accurately.
Although information culling at the quantum level may resemble biological
death, the later is a different process altogether and should not be confused
with the former process.

\section{Evolution and Correlation in Species}

In living systems, another important property is the interdependency between
different species. Different species could arise from mutation of some copies
in the natural process of evolution of the quantum species. Here we
discuss the question of evolution of a collection of species and if there
can be strong correlations between one of the mutated copies and the rest.
We are working within a model where there are only identical replicas
at our disposal in a closed universe.
If there is some interaction, it is possible that one of the copies can undergo
evolution leading to mutation described by
$\ver \psi \ra  \ver \psi \ra \rightarrow
\ver \psi \ra  U \ver \psi \ra=\ver \psi \ra \ver \psi' \ra$ or
it could be $\ver \psi \ra \ver \psi \ra \rightarrow
U \ver \psi \ra  \ver \psi \ra =\ver \psi' \ra \ver \psi \ra$,
where $\ver \psi' \ra = U|\psi\ra $ is the mutated copy.

Now we are looking for an interaction between two copies leading to
the evolution of one, yet
unknowable to us which one has actually undergone mutation.
This means that the first copy could have evolved or the second copy could
have. In other words, one of the evolved copies has been entangled with the
other copy. Specifically, we define
the process to be a unitary evolution that takes two copies of a living
organism $\ver \psi \ra  \ver \psi \ra$ and transforms them as

\begin{equation}
\ver \psi \ra  \ver \psi \ra \rightarrow N(\psi)
[\ver \psi \ra  U \ver \psi \ra + U \ver \psi \ra 
\ver \psi \ra],
\end{equation}
where $N(\psi)=1/\sqrt{2(1+\ver \la \psi \ver U \ver \psi \ra \ver^2)}$ is
the normalisation constant and $U$ is the local unitary operator that causes
evolution of one of the species. Can such an indistinguishable evolution of one
of the species occur in nature?

We will show that in a closed universe if there are an infinite number of
replicas, then there is no interaction between them that can cause an
evolution of one and entanglement between the evolved copy and the rest in
a deterministic manner.

\subsection{A paradox of mutation and entanglement}

We will argue that by entangling one
of the mutated copies with the rest of the species, we are increasing
the overlap between the initial and final living states (compared to
what would be the overlap between the
initial and the final unentangled states). It turns out that if we start with
$M$ copies of a quantum species, then in the limit of infinite number of copies
(where information has essentially become classical), the overlap between
the initial and the final states become one. This implies that the effect
(namely, the entangling of evolved copy with other copies) nullifies its cause
(namely, the interaction). But how can any effect in nature
nullify its cause? This is a paradox that we resolve in the folowing.

To see this interesting effect let us start with $M$ copies of a quantum
species and suppose that we are able to switch on an interaction between
them such that we have the following transformation
\begin{eqnarray}
\ver \Psi_i \ra =\ver \psi \ra^{\otimes M} &\rightarrow& N^{(M)}(\psi)
[\ver \psi \ra \ver \psi \ra  \cdots  U \ver \psi \ra
+ \ver \psi \ra \ver \psi \ra  \cdots
U \ver \psi \ra  \ver \psi \ra
\nonumber\\
& +&  \ldots + U \ver \psi \ra 
\ver \psi \ra \cdots \ver \psi \ra] = \ver \Psi_f^{\rm E} \ra,
\end{eqnarray}
where $N^{(M)}(\psi)=1/\sqrt{M+ M(M-1) \ver \la \psi \ver U \ver \psi
\ra \ver^2)}$
is the normalisation constant and $U$ is the unitary evolution that causes
mutation of one of the copy. If such an evolution is possible, then the
overlap between the initial and final entangled states is given by

\begin{eqnarray}
\ver \la \Psi_i \ver \Psi_f^{\rm E} \ra \ver^2 = \frac{M
\ver \la \psi \ver U\ver \psi \ra \ver^2 }{1 +
(M-1) \ver \la \psi \ver U \ver \psi \ra \ver^2}.
\end{eqnarray}

However, if there is no entanglement between the mutated copy and the rest,
then any one of the species will evolve in time and we will have

\begin{eqnarray}
\ver \Psi_i \ra =\ver \psi \ra^{\otimes M} \rightarrow
\ver \psi \ra  \ver \psi \ra \cdots  U \ver \psi \ra
= \ver \Psi_f^{\rm UE} \ra.
\end{eqnarray}
In the above we have assumed that the last copy evolves, but the argument holds
if any other copy evolves in time.
Therefore, the overlap between the initial and the final state is given by
$\ver \la \Psi_i \ver \Psi_f^{\rm UE} \ra \ver^2 =
\ver \la \psi \ver U\ver \psi \ra \ver^2$.
Now for any $M \ge 2$, we have
$\ver \la \Psi_i \ver \Psi_f^{\rm E} \ra \ver^2 \ge
\ver \la \Psi_i \ver \Psi_f^{\rm UE} \ra \ver^2$,
with equality sign holding when $\ver \la \psi \ver U\ver \psi \ra \ver=0$.
This shows that after the interaction, the entangling process leads
to increase of the overlap between the initial and the final state.
The crucial factor that make these states more indistinguishable is
$\frac{M}{1 + (M-1) \ver \la \psi \ver U \ver \psi \ra \ver^2}$ which is
greater than unity for $M\ge 2$. Interestingly, if we have infinite number
of copies of a living species, then the overlap becomes

\begin{eqnarray}
{\rm lim}_{M \rightarrow \infty} \ver \la \Psi_i \ver \Psi_f^{\rm E} \ra
\ver^2 =
{\rm lim}_{M \rightarrow \infty} \frac{M \ver \la \psi \ver U\ver \psi \ra \ver^2 }
{1 + (M-1) \ver \la \psi \ver U \ver \psi \ra \ver^2} \rightarrow 1
\end{eqnarray}
for all admissible $\ver \psi \ra$ and $U$.
This shows that if we start with an ensemble of identical prepared quantum
species, then the interaction leading to entanglement between one of the
mutated copies and the rest almost ceases the evolution, i.e. as if the system
has not evolved at all. It is an effect that nullifies its cause.
So what prohibits such a strange phenomenon in the quantum world?

\subsection{Impossibility of entangling a species with its
mutated replica}

The explanation for this paradox lies in the fact that a mutated
copy cannot be entangled with the rest of the species in the quantum
mechanical sense. We can illustrate this limitation for two copies of a
quantum species and then it holds for any number of copies. If the above
process exists for two arbitrary
non-orthogonal states, then for two copies of the quantum systems in the
states $\ver \psi \ra$ and $\ver \phi \ra$, we have

\begin{eqnarray}
\ver \psi \ra \ver \psi \ra &\rightarrow& N(\psi)
[\ver \psi \ra  U \ver \psi \ra + U \ver \psi \ra 
\ver \psi \ra], \nonumber\\
\ver \phi \ra  \ver \phi \ra &\rightarrow& N(\phi)
[\ver \phi \ra  U \ver \phi \ra  + U \ver \phi \ra 
\ver \phi \ra],
\end{eqnarray}
where $N(\phi)=1/\sqrt{2(1+\ver \la \phi \ver U \ver \phi \ra \ver^2)}$ is
the normalisation constant and $U$ is the same as before.
If such a quantum
mechanical process exists, then the unitarity of the interaction would mean
we must have the following requirement

\begin{equation}
\la \psi \ver \phi \ra^2 = 2 N(\psi) N(\phi)
[\la \psi \ver \phi \ra^2 + \la \psi \ver U \ver\phi \ra
\la \psi \ver U^{\dagger} \ver \phi \ra ].
\end{equation}
However, the above equation cannot hold for two non-orthogonal quantum species.
Hence, there can be no physical interaction (i.e. unitary process) that can
entangle one of the mutated copies with the original one in a deterministic
manner.

What is more surprising is that even two orthogonal species cannot
satisfy the above condition. When $\la \psi \ver \phi \ra=0$, still there
are two non-zero terms $\la \psi \ver U \ver\phi \ra$ and
$\la \psi \ver U^{\dagger}\ver \phi \ra$. To illustrate this point better,
consider two orthogonal states of a species. Let $\ver \psi \ra=\ver 0
\ra$ and $\ver \phi \ra=\ver 1 \ra$, then we have the interaction between
copies leading to

\begin{eqnarray}
\ver 0 \ra \ver 0 \ra & \rightarrow & N(0)
[\ver 0 \ra U \ver 0 \ra + U \ver 0 \ra  \ver 0 \ra] \nonumber\\
\ver 1 \ra \ver 1 \ra & \rightarrow & N(1)
[\ver 1 \ra U \ver 1 \ra + U \ver 1 \ra  \ver 1 \ra]
\end{eqnarray}
where we have taken $U\ver 0 \ra= a\ver 0 \ra+ b\ver 1 \ra$ and
$U\ver 1 \ra= a^* \ver 1 \ra - b^* \ver 0 \ra$. In this case
$\la \psi \ver U \ver\phi \ra \la \psi \ver U^{\dagger} \ver \phi \ra =
-{b^*}^2$ and hence (25) cannot be satisfied.

Since species in orthogonal states carry classical information, this shows
that classical information cannot be entangled with its mutated counterpart in
a deterministic manner. However, if we allow an ancillary system and
a non-unitary operation then it will be possible to satisfy (23) and (25) in a
probabilistic manner.


\section{Conclusion}

In this paper we have addressed various
quantum mechanical processes which are basic to life but we limit
ourselves to an `artifical living system' with possible implication to issues
in biological systems. We all know that real biological systems
(as with any meso or macroscopic systems) are in contact
with the environment, thus causing decoherence. Because of this contact, such
systems can therefore
show order even though the overall disorder of the Universe increases. In
other words, they do not violate Second Law of Thermodynamics since they are
just a subsystem. With a suitable amount of interaction with the outside
world (via exchange of energy, negentropy)
life can possibly emerge. One may argue that apparent order can possibly
emerge from Second Law, so there is no real mystery in that sense, when we 
realize
that overall disorder (entropy) is increasing. However, here we have analysed
such biological effects from the perspective of a closed quantum system.
Therefore, we do not enter
into the discussion of Second Law, nor exchange of energy with the
environment, which would surely play an important role. Future investigation
can throw more light on the question of replication, culling and
correlation with reference to Second Law of Thermodynamics when one takes
into account non-unitary evolutions in an open universe.

Merging of ideas from information theory and quantum theory has brought a
revolution and opened up a new field called quantum information science
\cite{nc}. Similarly, ideas from biology and quantum information may lead
to a new discipline called quantum bio-information theory.
Here, we have tried to bring out a physicist's approach to the problem
of life and some of its properties. Schr{\"o}dinger, one of the founders
of quantum mechanics, had initiated the question on how physical and chemical
laws can account for orderly and lawful events within a living organism. We
discussed Wigner's replication machine for a living state in the quantum
mechanical sense. However, it should be remarked that we have not dealt with
real biological systems in their full detail, rather we have worked within
a model of an `artificial life system' where species are described by quantum
mechanical amplitudes. Nevertheless, one can make some contact with `real'
biological systems. We elaborated the `no-cloning' and the `no-culling'
principles in the language of quantum bio-information. Impossibility of
the replication and culling of
quantum organism provide `permanence' to the information content.
We have shown that there cannot be any physical
interaction that can entangle a mutated quantum copy with the rest in a
deterministic manner. It is hoped that these discussions will be interesting
to those who wish to understand artificial or real `life' from a quantum
information theoretic angle and could have some implication in
quantum biology.\\


{\bf Acknowledgment:} I wish to thank C.~Fuch for bringing the paper
of E.~P.~Wigner which has been a source of inspiration. Also I thank
D.~Abbott for his encouragement throughout this project.


\end{document}